\documentclass[a4paper]{article}

\usepackage{INTERSPEECH2021,enumitem,bm}
\usepackage{amsmath}

\setlist[itemize]{align=parleft,left=0pt..1em}

\DeclareMathOperator{\similarity}{sim}

\title{Emotion Recognition from Speech Using Wav2vec 2.0 Embeddings}
\name{Leonardo Pepino$^{1,2}$, Pablo Riera$^1$, Luciana Ferrer$^1$}
\address{\vspace{-0.15cm}
  $^1$Instituto de Investigaci\'on en Ciencias de la Computaci\'on (ICC), CONICET-UBA, Argentina
  $^2$Departamento de Computaci\'on, FCEyN, Universidad de Buenos Aires (UBA), Argentina}
\email{lpepino@dc.uba.ar, priera@dc.uba.ar,lferrer@dc.uba.ar}

\begin{document}

\maketitle
\begin{abstract}
Emotion recognition datasets are relatively small, making the use of the more sophisticated deep learning approaches challenging. In this work, we propose a transfer learning method for speech emotion recognition where features extracted from pre-trained wav2vec 2.0 models are modeled using simple neural networks. We propose to combine the output of several layers from the pre-trained model using trainable weights which are learned jointly with the downstream model. Further, we compare performance using two different wav2vec 2.0 models, with and without finetuning for speech recognition.
We evaluate our proposed approaches on two standard emotion databases IEMOCAP and RAVDESS, showing superior performance compared to results in the literature. 
\end{abstract}
\noindent\textbf{Index Terms}: speech emotion recognition, transfer learning, wav2vec 2.0

\section{Introduction}

In recent years, progressively more people are interacting with virtual voice assistants, such as Siri, Alexa, Cortana and Google Assistant. Interfaces that ignore a user’s emotional state or fail to manifest the appropriate emotion can dramatically impede performance and risks being perceived as cold, socially inept, untrustworthy, and incompetent \cite{brave2009emotion}. Because of this, speech emotion recognition is becoming an increasingly relevant task.

Several datasets have been created over the years for training and evaluating emotion recognition models, including SAVEE \cite{savee}, RAVDESS \cite{ravdess}, EMODB \cite{emodb}, IEMOCAP \cite{iemocap}, and MSP-Podcast \cite{lotfian-msppodcast-2019}. With the exception of MSP-Podcast, these datasets are relatively small in size, usually including only a few dozen speakers. 
In terms of modeling techniques, many different traditional approaches have been proposed, using hidden Markov models (HMMs) \cite {schuller-hmm-2003, sato-hmm-2007}, support vector machines (SVMs) \cite {shen-svm-2011,rao-svm-2013},  decision trees \cite {borchert-dt-2005, lee-dt-2011}, and, more recently, deep neural networks (DNNs) \cite {mirsamadi-rnn-2017, satt-cnnlstm-2017, sarma-raw-2018}. Yet, while DNNs have shown large gains over traditional approaches on tasks like automatic speech recognition (ASR) \cite{gulati-conformer-2020} and speaker identification \cite{matvejka-speakerid-2016}, the gains observed on emotion recognition are limited, likely due to the small size of the datasets.

A common strategy when dealing with small training datasets is to apply transfer learning techniques.
One approach for transfer learning is to use a model learned for a certain auxiliary task for which large datasets are available for training to improve robustness for the task of interest for which data is scarce. The model learned on the auxiliary task can be used as feature extractor or fine-tuned, after replacing some of its final layers, to the task of interest.
Recently, transfer learning approaches have been explored in the field of speech emotion recognition. In \cite{zhao-tl-2021}, a deep neural network based on transformers \cite{vaswani-attention-2017} is pretrained on LibriSpeech \cite{librispeech} using multiple self-supervised objectives at different time scales. Then, the model is used as a feature extractor or fine-tuned for speech emotion recognition, among other downstream tasks. Similarly, in \cite{li-tl-2021} and \cite{zhang-tl-2021}, a deep encoder is pretrained in a contrastive predictive coding task, and the resulting embeddings are tested in speech emotion datasets.
Recently, several models for automatic speech recognition (ASR) which use self-supervised pretraining have been released, including wav2vec \cite{wav2vec} and VQ-wav2vec \cite{vq-wav2vec}. A few recent studies \cite{siriwardhana-vqw2v-2020, macary-w2v-2020, boigne-w2v-2020} have successfully applied representations from these models as features for  emotion recognition.

In line with those works, in this paper, we explore the use of the wav2vec 2.0 model \cite{wav2vec2}, an improved version of the original wav2vec model, as a feature extractor for speech emotion recognition. 
The main contributions of our paper are (1) the use of wav2vec 2.0 representations for speech emotion recognition which, to our knowledge, had never been done for this task, (2) a novel approach for the downstream model which leverages information from multiple layers of the wav2vec 2.0 model and leads to significant improvements over previous approaches, and (3) an analysis of the importance of the different layers in wav2vec 2.0 for the emotion recognition task. Our results are superior to others in the literature for models based only on acoustic information for IEMOCAP and RAVDESS. The code to replicate the results of this paper will soon be released at https://github.com/habla-liaa/ser-with-w2v2.

\section{Methods}
In our study, we extracted features from two released wav2vec 2.0 models and used them for speech emotion recognition. In this section, we describe the wav2vec 2.0 model, the datasets used for training and evaluation, and the downstream models.

\vspace{-0.1cm}\subsection{Wav2vec 2.0 model architecture}
\vspace{-0.1cm}
Wav2vec 2.0 \cite{wav2vec2} is a framework for self-supervised learning of representations from raw audio. The model consists of three stages. The first stage is a {\bf local encoder}, which contains several convolutional blocks and encodes the raw audio into a sequence of embeddings with a stride of 20 ms and a receptive field of 25 ms. Two models have been released for public use, a large one and a base one where the embeddings are 1024- and 768-dimensional, respectively. The second stage is a {\bf contextualized encoder}, which takes the local encoder representations as input. Its architecture consists of several transformer encoder blocks \cite{vaswani-attention-2017}. The base model uses 12 transformer blocks with 8 attention heads each, while the large model uses 24 transformer blocks with 16 attention heads each. Finally, a {\bf quantization module}, takes the local encoder representations as input and consists of 2 codebooks with 320 entries each. A linear map is used to turn the local encoder representations into logits. Given the logits, Gumbel-Softmax \cite{jang-gumbelsoftmax-2016} is applied to sample from each codebook. The selected codes are concatenated and a linear transformation is applied to the resulting vector leading to a quantized representation of the local encoder output, which is used in the objective function, as explained below.

\subsection{Wav2Vec 2.0 pretraining and finetuning}
\vspace{-0.1cm}
The wav2vec 2.0 model is pretrained in a self-supervised setting, similar to the masked language modelling used in BERT \cite{devlin-bert-2018} for NLP. Contiguous time steps from the local encoder representations are randomly masked and the model is trained to reproduce the quantized local encoder representations for those masked frames at the output of the contextualized encoder.
 
The training objective is composed by terms of the form 
\begin{equation}
    L_m = -\log \frac{\exp(\similarity(c_t,q_t)/\kappa)}{\sum_{\tilde{q} \in \tilde{Q}}{\exp(\similarity(c_t,\tilde{q})/\kappa)}} 
\label{eq:wav2vec2_loss}
\end{equation}
where $\similarity(c_t,q_t)$ is the cosine distance between the contextualized encoder outputs $c_t$ and the quantized local encoder representations $q_t$. $t$ is the time step,  $\kappa$ is the temperature and \~{Q} is the union of a set of K \emph{distractors} and $q_t$. The distractors are outputs of the local encoder sampled from masked frames belonging to the same utterance as $q_t$. The contrastive loss is then given by $L_m$ summed over all masked frames. Finally,  terms to encourage diversity of the codebooks and L2 regularization are added to the contrastive loss.  

The main goal of the wav2vec 2.0 paper was to use the learned representations to improve ASR performance, requiring less data for training and enabling its use for low resource languages.
To this end, the model trained as described above is finetuned for ASR using a labelled speech corpus like LibriSpeech. A randomly initialized linear projection is added at the output of the contextual encoder and the connectionist temporal classification (CTC) loss \cite{graves-ctc-2006} is minimized. The models finetuned in 960 hours of LibriSpeech reach state of the art results in automatic speech recognition when evaluated in different subsets of LibriSpeech. Even when finetuning using considerably less hours, wav2vec 2.0 models reach a performance comparable to the state of the art.

In this paper we compared the performance in speech emotion recognition when using both the wav2vec 2.0 base model pretrained in Librispeech without finetuning\footnote{https://dl.fbaipublicfiles.com/fairseq/wav2vec/wav2vec\_small.pt} (we will call this model Wav2vec2-PT), and a model finetuned for ASR using  a 960-hour subset of Librispeech\footnote{https://dl.fbaipublicfiles.com/fairseq/wav2vec/wav2vec\_small\_960h.pt} (Wav2vec2-FT). In both cases, we used the base model as we did not see significant performance improvements when using the large one, and it allowed us to reduce computational requirements.

\begin{figure}[t]
    \centering
    \includegraphics[width=0.9\linewidth]{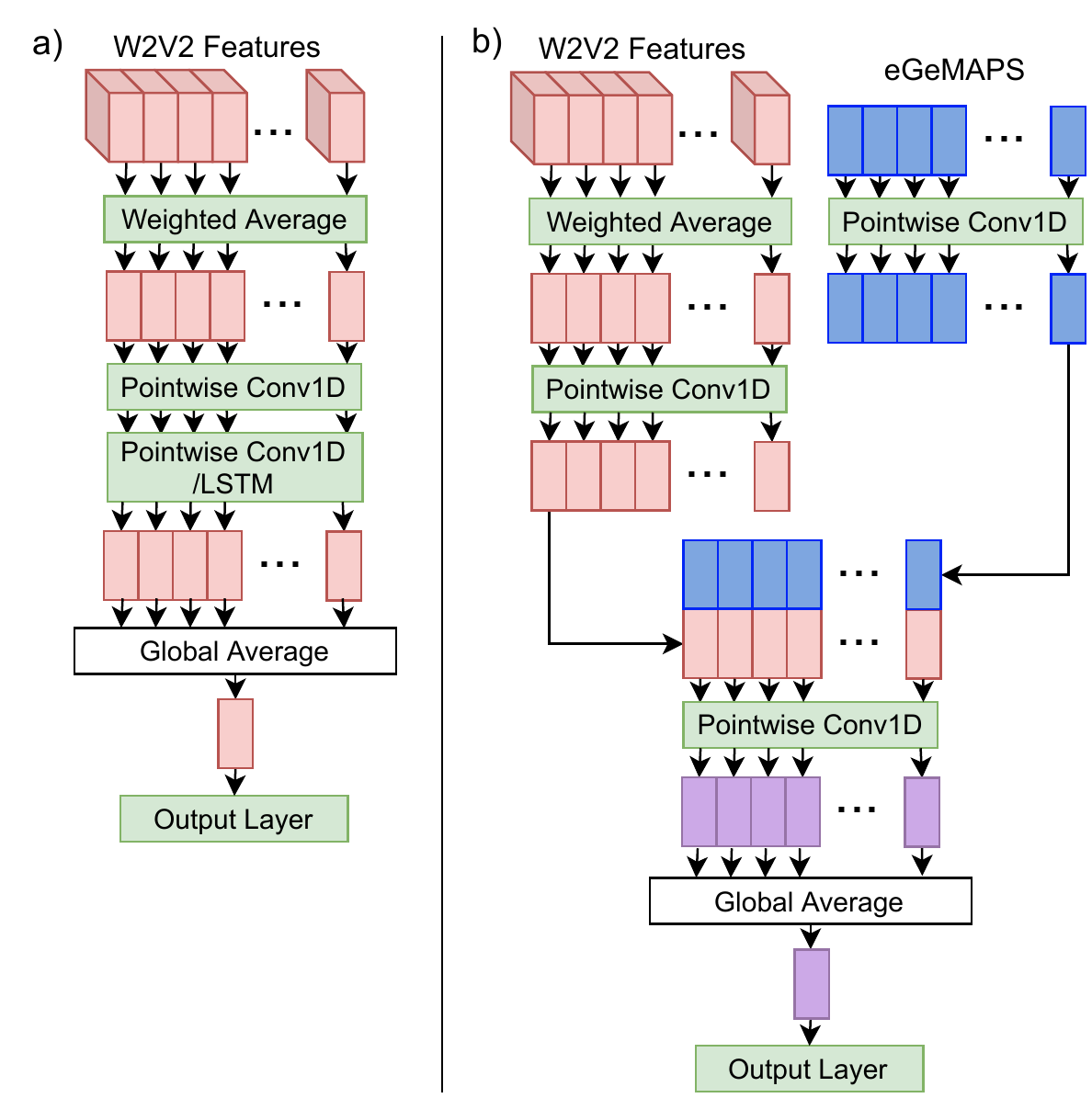}
    \vspace{-0.1cm}
    \caption{Downstream models used for this work. The green blocks represent trainable layers. We call the model in figure a) Dense, when using a pointwise convolutional layer, and LSTM, when using an LSTM layer. We call the model in b) the Fusion model. Blocks for each time-step have a height dimension for the features, and an optional depth dimension for the layers in the wav2vec 2.0 model.     }
    \label{fig:downstream-models}
\end{figure}

\vspace{-0.1cm}
\subsection{Features}
\vspace{-0.1cm}
We compare results when using the output of the local and the contextualized encoders as input to the downstream models. We also propose to use a weighted average of the outputs of the 12 transformer blocks, along with those of the local encoder. The weights of this average are trained along with the rest of the downstream model, as explained in the next subsection.

As baseline features, we also calculated magnitude spectrograms with a hanning window of 25 ms and a hop size of 10 ms, and eGeMAPS \cite{eyben-geneva-2015} low level descriptors (LLD), which are commonly used in the emotion recognition literature. The eGeMAPS features were extracted using opensmile-python~\cite{eyben-opensmile-2010}. In order to match the sequence lengths of the eGeMAPS features, which use a stride of 10 ms, with the lengths of the wav2vec 2.0 features, which have a stride of 20 ms, we downsampled the eGeMAPS LLDs by averaging every 2 consecutive frames.

All features were normalized by subtracting the mean and dividing by the standard deviation of that feature over all the data for the corresponding speaker. When disabling speaker normalization for comparison of results, we replaced it with global normalization per feature. In this case, the statistics are computed over the training data only, excluding the test data.

\subsection{Downstream models}
\vspace{-0.1cm}
Our downstream model is inspired by \cite{boigne-w2v-2020} due to its simplicity, which should reduce the chance of overfitting. 
The model we will refer to as \emph{Dense}, consists of 2 dense layers with 128 neurons, ReLU activation and dropout with a probability of 0.2, applied independently to each frame of the input features sequence~$f$. This is equivalent to using 1D pointwise convolutional layers (with a kernel size of 1) and 128 filters. The outputs are averaged over time resulting in a vector of size 128. Finally, another dense layer with softmax activation returns probabilities for each of the emotion classes. During the training of the downstream model, the weights of the wav2vec 2.0 model remain unaltered, so it serves as a feature extractor. 

For computational reasons, we used a maximum sequence length of 400 for IEMOCAP and 250 for RAVDESS, as input to the network. This is equivalent to 8 seconds and 5 seconds, respectively. Note that the output of the contextualized encoder can contain information from the full input waveform which might be longer than the sequence seen by the network. This is because each output of the contextual encoder has a receptive field given by the whole input waveform, since this encoder is a transformer. On the other hand, the local encoder has a limited receptive field, so its outputs can only capture local information. 

For the case in which we take as features the activations from both the wav2vec 2.0 local encoder and the transformer blocks, we incorporate a trainable weighted average as the first layer. This layer learns the weights $\alpha_i$ for each of the wav2vec 2.0 layer activations, $f_i$, where $f_0$ corresponds to the local encoder output, $f_1$ through $f_{11}$ are the outputs of the internal blocks in the transformer and $f_{12}$ is the output of the contextualized encoder. Then, the activations are combined as follows
\begin{equation}
    f = \frac{\sum_{i=1}^{N}{\alpha_i f_i}}{\sum_{i=1}^{N}{\alpha_i}}
\end{equation}
The weights $\alpha_i$ are initialized with 1.0.
Note that the layer activations can be combined in this way because they are all the same size.
This way of extracting features from a pretrained model is similar to the approach used in ELMO \cite{peters-elmo-2018} for NLP.

In the results section, we compare performance of the Dense model with the LSTM model, in which the second dense layer is replaced by an LSTM layer. Finally, we also evaluate a third model, called Fusion model, which incorporates a branch taking as input eGeMAPS features. In this last model, the outputs of the first dense layer are concatenated before applying the second dense layer. The described downstream models can be seen in Figure \ref{fig:downstream-models}.

The downstream models were trained using batches of 32 utterances, Adam optimizer with a learning rate of 0.001, and early stopping with a patience of 4 epochs monitoring the validation loss.

\subsection{Datasets}
\vspace{-0.1cm}
The Ryerson Audio-Visual Database of Emotional Speech and Song
(RAVDESS) \cite{ravdess} is a  multi-modal database of
emotional speech and song. It features 24 different actors (12 males and 12 females) enacting 2 statements: “Kids are
talking by the door” and “Dogs are sitting by the door.” with 8 different emotions: happy, sad, angry, fearful, surprise,
disgust, calm, and neutral. These emotions are expressed in two different intensities: normal and strong, except for neutral (normal only). Each of the combinations was spoken and sung, and repeated 2 times, leading to 104 unique vocalizations per actor.
Following \cite{venkataramanan2019emotion}, we merged the neutral and calm emotions, resulting in 7 emotions, and used the first 20 actors for training, actors 20-22 for validation to do early stopping, and actors 22-24 for test.

The  Interactive  Emotional  Dyadic  Motion  Capture
(IEMOCAP) dataset \cite{iemocap} has a length of approximately 12 hours and consists of scripted and improvised dialogues by 10 speakers. It is composed of 5 sessions, each including speech from an actor and an actress. Annotators were asked to label each sample choosing one or more labels from a pool of emotions. In this work, we used 4 emotional classes: anger, happiness, sadness and neutral, and following the work in \cite{Fayek2017},  we relabeled excitement samples as happiness. Instances from other classes and with no majority label across the annotations were discarded.\footnote{Note that discarding no-agreement samples and samples from non-target emotions is not an ideal practice \cite{riera2019}. Here, we decided to do this since it is standard practice in emotion recognition literature, facilitating comparisons across papers.} We also trimmed the waveforms to 15 seconds to reduce the computational requirements when extracting wav2vec 2.0 features. Only 2\% of the waveforms were longer than 15 seconds.
To evaluate the models in IEMOCAP, we performed 5-fold cross-validation, leaving one session out for each of the folds, as it is a standard practice with this dataset. We used one of the 4 training sessions to perform early stopping for each cross-validation model. 

\begin{table}[t]
    \centering
    \caption{Average recall (\%) obtained for the different features and models explored in IEMOCAP and RAVDESS. All results are obtained with the Dense downstream model.}
    \begin{tabular}{llcc}
    \toprule
          Pretrained  & Features & \multicolumn{2}{c}{Dataset} \\
     model &  & IEMOCAP & RAVDESS\\
    \midrule
     None & eGeMAPS&     52.4 $\pm$ 0.1 &     57.0 $\pm$ 2.4\\
      & Spectrogram&     49.8 $\pm$ 1.0 &     44.5 $\pm$ 0.8\\
            \midrule
    Wav2vec2-PT & Local enc. &     60.3 $\pm$ 0.7 &     65.4 $\pm$ 1.7\\
                & Cont. enc. &     58.5 $\pm$ 0.6&     69.0 $\pm$ 0.2\\
                & All layers &     \textbf{67.2} \boldsymbol{$\pm$} \textbf{0.7} &     \textbf{84.3}  \boldsymbol{$\pm$} \textbf{1.7}\\
            \midrule
    Wav2vec2-FT & Local enc. &     57.3 $\pm$ 1.0&     58.8 $\pm$ 2.7\\
                   & Cont. enc. &     44.6 $\pm$ 1.0&     37.5 $\pm$ 3.0\\
                   & All layers &     63.8 $\pm$ 0.3&     68.7 $\pm$ 0.9\\
    \bottomrule
    \end{tabular}
    \label{tab:results-features}
\end{table}

\section{Results and discussion}

We calculated the performance of our models by training them 5 times with different seeds. Table \ref{tab:results-features} shows the average recall over all emotion classes obtained using the different features extracted from wav2vec 2.0 models. We also show two systems based on eGeMAPS and spectrogram features, which can be considered as baselines. In all cases, features are normalized by speaker and the Dense model architecture in Figure \ref{fig:downstream-models} is used as downstream model. We can see that features for both wav2vec 2.0 models, the one finetuned in 960 hours of Librispeech (wav2vec2-FT) and the one that is not finetuned (wav2vec2-PT), the local encoder representations lead to better results than both of the baseline features. It is worth noting that eGeMAPS, spectrograms and the wav2vec 2.0 local encoder representations contain information restricted only to a local window around each frame, of 60 ms for eGeMAPS, and 25 ms for the others. Also, the downstream model combines information from consecutive frames using just global average, which is a very simple approach that might be suboptimal because it cannot take into account temporal patterns in the features. In spite of that, the local encoder representations, particularly the ones obtained from the PT model, reach a performance comparable to much more complex models like the one proposed in \cite{satt-cnnlstm-2017}, which is a CNN with Bi-LSTM layers trained in a fully supervised setting with spectrograms as input.

We can also see that the wav2vec2-PT features perform better than wav2vec2-FT features in all cases. In particular, the model using only the contextualized encoder outputs for the wav2vec2-FT model has the worst results in the table. We hypothesize that this is because when the model is finetuned for an ASR task, information that is not relevant for that task but might be relevant for speech emotion recognition is lost from the embeddings. For example, information about the pitch might not be important for speech recognition, while it is essential for speech emotion recognition. 

Further, Table \ref{tab:results-features} shows that using a weighted average of the transformer blocks outputs along with the local encoder outputs (rows labelled ''All layers'') results in better performance than using only the local or the contextual encoder outputs. Using only the local encoder representations might not give information about events occurring at time scales larger than the receptive field (25 ms). Further, the output of the contextual encoder might be similar to the quantized local encoder representations, since the contrastive loss objective is designed to achieve this goal. Hence, using both of these layers, along with all intermediate layers in the transformer provides additional information that is valuable to the model.

\begin{figure}[t]
    \centering
    \includegraphics[width=\linewidth]{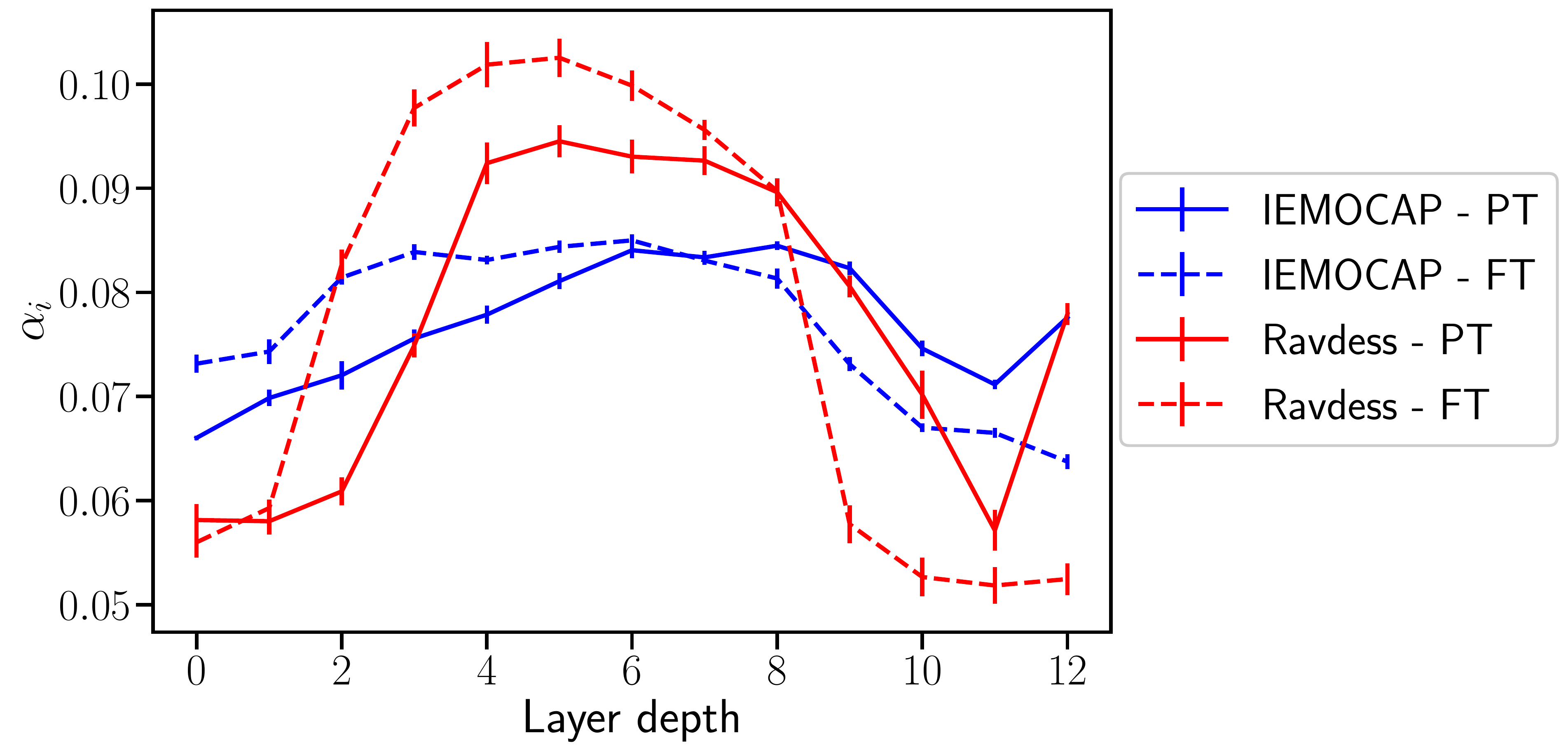}
    \vspace{-0.5cm}
    \caption{Weights of the trainable weighted average layer of the dense model. Index 0 corresponds to the local encoder output, and indices 1 through 12 correspond to the outputs of the transformer blocks in the contextual encoder, with layer 12 being the output layer. Solid lines correspond to the wav2vec2-PT features, while dashed lines correspond to the wav2vec2-FT features.
    }
    \label{fig:layerweights}
\end{figure}

Figure \ref{fig:layerweights} shows the weights $\alpha_i$ from the weighted average layer once the downstream models are trained.
We can see that the middle layers are given larger weights. This could be because these layers have already contextualized enough information from the local encoder, but they are not yet too specific to the pretraining or finetuning task as the last layers. Also, in the case of the feature extracted from the wav2vec2-PT model, the weights tend to be larger in the layers closer to the output when compared with the weights for the features from the wav2vec2-FT model. This again suggests that the layers of the wav2vec2-FT model closer to the output are less useful for emotion recognition than those of the wav2vec2-PT model.
Finally, note that in both datasets, the different training seeds lead to very similar weights, as observed from the error bars in Figure \ref{fig:layerweights}, from which we can conclude that these weights are not too sensitive to the neural network initialization.

Finally, we experimented with several variations of the best performing model in Table~\ref{tab:results-features}. The first line in Table~\ref{tab:variant_results} shows the results for that system, corresponding to the fifth line in Table~\ref{tab:results-features}. 
The rest of the lines in Table~\ref{tab:variant_results} show results for that model and the others in Figure \ref{fig:downstream-models} applying global normalization instead of speaker-dependent normalization. 
This is the scenario that is most commonly used in papers as is the one with less assumptions about the available data, treating all samples from a speaker as independent from each other, not assuming that additional samples from a speaker are available at test time. 

Comparing the first and the second line in Table \ref{tab:variant_results}, we can see that results significantly degrade when doing global normalization. This is expected since normalization by speaker helps the model focus on the emotion characteristics by eliminating part of the speaker information. The degradation is larger in RAVDESS probably because, in this dataset, audios from different emotions do not have much variation in lexical content. Hence, by reducing the effect of speaker in the features all that is left is the variation due to the emotions. On the other hand, in IEMOCAP data, the variation due to lexical content is still in the samples after speaker normalization.

The third and fourth lines in Table \ref{tab:variant_results} show the results obtained using the LSTM model, and the Fusion model. The latter fuses eGeMAPS with the wav2vec 2.0 features. 
We can see that using an LSTM layer before the global pooling, does not seem to bring improvements over using a simple dense layer. This might be because wav2vec 2.0 features are already contextualized and have global information about the full utterance. Also, LSTMs might be more prone to overfitting or optimization problems than a simpler dense layer. Finally, the table shows some modest improvements from the addition of eGeMAPS features, suggesting that wav2vec 2.0 features may be lacking some of the information present in eGeMAPS. 

\begin{table}[t]
    \centering
    \caption{Average recall from the model variants in Figure \ref{fig:downstream-models} using the wav2vec2-PT features from all layers, with two types of normalization (top block) compared with results from other works, which all use global normalization (bottom block). In bold are the best results using global normalization.}
    \begin{tabular}{lcc}
    \toprule
    Model - Norm & IEMOCAP & RAVDESS \\
    \midrule
    Dense - Speaker & 67.2 $\pm$ 0.7 & 84.3 $\pm$ 1.7 \\
    Dense - Global &   65.8 $\pm$ 0.3 &  75.7 $\pm$ 2.3 \\
    LSTM - Global  &   64.8 $\pm$ 1.9 &   74.6 $\pm$ 3.7 \\
    Fusion - Global &   \textbf{66.3} \boldsymbol{$\pm$} \textbf{0.7}&   \textbf{77.5} \boldsymbol{$\pm$} \textbf{1.0}\\
    \midrule
    BiLSTM w. attention \cite{mirsamadi-rnn-2017} & 58.8 & - \\
    CNN-BiLSTM \cite{satt-cnnlstm-2017} & 59.4 & - \\
    TDNN-LSTM w. attention \cite{sarma-raw-2018} & 60.7 & - \\
    Wav2Vec \cite{boigne-w2v-2020} & 64.3 & - \\
    \bottomrule
    \end{tabular}
    \label{tab:variant_results}
\end{table}

Table \ref{tab:variant_results} also shows some of the results obtained in other works on the IEMOCAP dataset, using the same experimental setup we are using. For a fair comparison with our work, we restrict the comparison to models in the literature that do not use automatic or manual transcriptions.
We can see that all of our models perform better than the state of the art. 

Finally, we also compare our results with those in \cite{issa-ser-2020}, which consists of a deep convolutional neural network using acoustic features as input. For this comparison, we used the dense downstream model with global normalization and imitated their experimental setup both for IEMOCAP and RAVDESS. For IEMOCAP, they only use the improvised sessions and full agreement utterances. For RAVDESS, they use 5 fold cross-validation, dividing the data randomly. Moreover, they do not merge calm and neutral emotions, so the total number of emotions to be predicted is 8. We outperformed the models in \cite{issa-ser-2020} for both datasets obtaining an average recall of 84.1 $\pm$ 1.2 \% in RAVDESS, and 72.1 $\pm$ 0.9\% in IEMOCAP, compared to the results in the paper which are  64.3\%
and 71.6\%, respectively.

\section{Conclusions}
In this work, we explored different ways of extracting and modeling features from pretrained wav2vec 2.0 models for speech emotion recognition. We proposed to combine the different layers in the wav2vec 2.0 model using trainable weights and model the resulting features with a simple DNN with a time-wise pooling layer. We evaluated our models on two standard emotion datasets, IEMOCAP and RAVDESS, and showed superior results on both cases, compared to those in recent literature. We found that the combination of information from different layers in the wav2vec 2.0 model led to improved results over using only the encoder outputs, as in previous works. 
Further, we found that the combination of the wav2vec 2.0 features with a set of prosodic features gave additional gains, suggesting that the wav2vec 2.0 model does not contain all the prosodic information needed for emotion recognition.
Finally, we showed that a wav2vec 2.0 model finetuned for the task of ASR worked worse than the one trained only with the self-supervised task, indicating that the finetuning eliminates information from the embeddings that is useful for emotion recognition.

\section{Acknowledgements}
This material is based upon work supported by a Google Faculty Research Award, 2019, and an Amazon Research Award, 2019.
\bibliographystyle{IEEEtran}

\bibliography{mybib}

\end{document}